\documentclass[12pt,preprint]{aastex}
\usepackage{lscape}

\newcommand{\bdv}[1]{\mbox{\boldmath$#1$}}

\def\rel{{\rm rel}}
\def\e{{\rm E}}
\def\au{{\rm AU}}
\def\muas{{\mu\rm as}}
\def\kms{{\rm km}\,{\rm s}^{-1}}

\def\pc{{\rm pc}}
\def\rel{{\rm rel}}
\def\max{{\rm max}}

\def\e{{\rm E}}
\def\bpi{{\bdv{\pi}}}
\def\bmu{{\bdv{\mu}}}
\def\btheta{{\bdv{\theta}}}

\begin{document}
\title{The Mass of the MACHO-LMC-5 Lens Star}

\author{Andrew Gould}
\affil{Department of Astronomy, The Ohio State University, 
140 West 18th Avenue, Columbus, OH 43210\\
gould@astronomy.ohio-state.edu
}
\author{David P. Bennett}
\affil{Department of Physics, Notre Dame University, Notre Dame, IN 46556\\
bennett@nd.edu}
\and
\author{David R. Alves}
\affil{Department of Astronomy, Columbia University, 550 West 120th St.,
New York, NY 10027\\
alves@astro.columbia.edu}

%\submitted{Submitted to The Astrophysical Journal}

\begin{abstract}
We combine the available astrometric and photometric data for the 
1993 microlensing event MACHO-LMC-5 to measure the mass of the lens,
$M=0.097\pm 0.016\,M_\odot$.  This is the most precise direct
mass measurement of a single star other than the Sun.  In principle,
the measurement error could be reduced as low as 
10\% by improving
the trig parallax measurement using, for example, the
{\it Space Interferometry Mission}.  Further improvements might
be possible by rereducing the original photometric lightcurve
using image subtraction or by obtaining new, higher-precision baseline
photometry of the source.  We show that the current data strongly
limit scenarios in which the lens is a dark (i.e., brown-dwarf) companion
to the observed M dwarf rather than being the M dwarf itself.
These results set the stage for a confrontation between mass
estimates of the M dwarf obtained from spectroscopic and photometric
measurements and a mass measurement derived directly from the
star's gravitational influence.  This would be the first such
confrontation for any isolated star other than the Sun.
\end{abstract}


\section{Introduction
\label{intro}}

Direct measurements of stellar masses provide an essential foundation for
theoretical models of stars.  Such measurements must be free of
model-dependent assumptions about the star's internal physics and so
can be obtained only from the star's gravitational effects on external 
objects.   Almost all stars with directly measured
masses are components of binaries.  However,
binary stars  may not always evolve 
as do single stars, and thus direct mass measurements of single stars are of 
prime importance.
The Sun's mass has been measured
using two completely independent methods: first by applying
Newton's generalization of Kepler's Third Law to the motion
its companions (the planets); second from its deflection
of light from distant stars \citep{eddington,fma}.  For
all stars other than the Sun, direct mass measurements
have, until recently, only been possible with the first method.
That is, direct mass measurements have been restricted to
components of binary systems with orbital periods of decades or
less.

All light-deflection methods rely fundamentally on the equation,
\begin{equation}
\theta_\e = \sqrt{\kappa M \pi_\rel},\quad 
\kappa\equiv {4 G\over \au\,c^2} \simeq {8.14\,{\rm mas}\over M_\odot},
\label{eqn:thetadef}
\end{equation}
where $\theta_\e$ is the angular Einstein radius, $M$ is the mass of
the lens, and $\pi_\rel$ is the lens-source relative parallax.
\citet{refsdal} was the first to propose that the masses of
stars could be measured from light deflection.  His method was
purely astrometric.  By measuring the relative separation of
a nearby lens and a more distant source, one could directly determine
the relative parallax $\pi_\rel$, the angular impact parameter $\beta$,
and the maximal deflection $\Delta\theta$.  In the simplest (and typical) case,
$\beta\gg \theta_\e$,
\begin{equation}
\theta_\e^2 = \beta\Delta\theta, \qquad (\beta\gg\theta_\e).
\label{eqn:widesep}
\end{equation}
Hence, combining equations~(\ref{eqn:thetadef}) and (\ref{eqn:widesep})
leads to a simple expression for the mass.  \citet{refsdal}'s method
will be carried out for perhaps a dozen nearby stars using the
{\it Space Interferometry Mission} ({\it SIM}, \citealt{sg00}).

A second method was proposed by \citet{gould92} and first carried
out by \citet{alcock95}.  Here one uses the
accelerated platform of the Earth to measure the microlens parallax,
\begin{equation}
\pi_\e = \sqrt{\pi_\rel\over\kappa M},
\label{eqn:piedef}
\end{equation}
and combines this with an independent measurement of $\theta_\e$ to
obtain the mass.  The method yields the mass and, simultaneously, the 
lens-source relative parallax $\pi_\rel$:
\begin{equation}
M={\theta_\e \over \kappa\pi_\e},\qquad
\pi_\rel = \pi_\e\theta_\e.
\label{eqn:masspirel}
\end{equation}
The major problem here is
that the number of events that last long enough for the Earth's
acceleration to significantly affect the lightcurve is quite
small, and for only a small minority of these events is it possible
to measure $\theta_\e$.  Indeed, there is only one event with a
precisely measured mass using this technique \citep{jin}, and
that event is a binary.  There are, in addition, two 
single-star events to which this method has been applied, although
the mass measurements are much cruder.  For one, OGLE-2003-BLG-238,
the mass is measured to only a factor of a few \citep{jiang}.
The other is MACHO-LMC-5, which is the subject of this paper.

In addition, \citet{ghosh} have shown that the mass of 
OGLE-2003-BLG-145/MOA-2003-BLG-45 could be determined precisely
provided that the lens-source relative proper motion is measured
astrometrically.
One variant of this method, proposed by \citet{delplancke01}, 
is to measure $\pi_\e$ using the accelerated platform of the Earth and 
$\theta_\e$ from the centroid
shift of the source during the event \citep{my95,hnp95,walker95}.  This method
would be especially applicable for black-hole microlensing candidates
\citep{bennett02,mao02,agol}, which are expected to have large enough
$\theta_\e$ to allow such measurement from the ground and which
are dark (and so ineligible for the relative proper-motion method of
\citealt{ghosh}).

Finally, this approach can be extended to a greater number of
shorter-duration events by measuring 
the microlens parallax $\pi_\e$
by comparing the photometric events as seen from the Earth and
a satellite in solar orbit \citep{refsdal66,gould95} and 
combining this with an astrometric measurement of $\theta_\e$ using a
space interferometer \citep{bsv,pac98}.  
Of order 200 such mass measurements will be made by {\it SIM} \citep{gs99},
which will itself be in solar orbit.

MACHO-LMC-5 is unique in that it has both a measurement
of $\bpi_\e$ derived from the microlensing lightcurve
\citep{alcock01,gould04}, to which we refer in this paper as a
``photometric'' quantity, and a completely independently determined
full post-event astrometric solution, including both
$\pi_\rel$ and the lens-source relative proper motion $\bmu_\rel$
\citep*{alcock01,drake}.
Here, $\bpi_\e$ is the vector microlens parallax, whose magnitude
is $\pi_\e$ and whose direction is that of the lens-source relative motion.
%That is,
%\begin{equation}
%{\bpi_\e\over \pi_e} = {\bmu_\rel\over \mu_\rel}.
%\label{eqn:bpiedef}
%\end{equation}

Thus, MACHO-LMC-5 is of particular interest for several reasons.
First, it permits three 
independent tests on the consistency of the measurements.
Second, if the measurements pass these consistency tests, they can be combined
to obtain a more accurate estimate of the mass.  Third, 
the very high
magnification of the microlensing event $A_\max\sim 80$ permits one
to place very strong constraints on the lens being a close binary rather
than a single star.  
Fourth, 
the photometric and
astrometric measurements can be combined to test the hypothesis that the 
resolved star that appears to be the lens in the astrometric images is not
in fact the lens but rather is merely
 a luminous binary companion to a dark substellar 
object that generated the microlensing event.  
Finally, the successful completion of all these
tests would allow a direct confrontation between the measured mass
of a single star and its mass as predicted from photometric and
spectroscopic observations.  This would be the first such confrontation
for any isolated star other than the Sun.

Here we build on the work of \citet{alcock01}, \citet{gould04}, and 
\citet{drake} to carry out the above-described 
tests, in so far as it is possible
today, and outline how these tests can be further refined in the
future.  The error in our mass estimate, 17\%, is the smallest for any
direct mass measurement of a single star other than the Sun, and
even approaches the precision of measurements of M-dwarf masses from
binaries \citep{delfosse}.

\section{Brief History
\label{sec:history}}

\citet{alcock01} originally measured $\bmu_\rel$ by analyzing epoch 1999 
{\it Hubble Space Telescope (HST)} Wide Field and Planetary Camera
(WFPC2) images obtained by
\citet{alcockprelim}.  To do so, they assumed that the two resolved objects
in those images were the lens and the source that had been virtually coincident
6.3 years earlier when the event occurred in 1993.  They checked for
consistency between the direction of $\bmu_\rel$ so obtained and
the direction of $\bpi_\e$ (measured in the heliocentric frame)
as determined from a microlens parallax analysis of the 
microlensing-event lightcurve.  They found that these position angles
agreed at the
$1\,\sigma$ level.  By combining the astrometric and lightcurve data,
they determined both $\pi_\e$ and $\theta_\e$ and so, using 
equation~(\ref{eqn:masspirel}), both the mass and distance of the lens.
The best estimate of the lens mass was substantially below the
hydrogen-burning limit (albeit with moderately large error bars) and
so appeared somewhat inconsistent with the hypothesis that the
luminous star seen in the images was actually the lens.  The inferred,
distance, $D_{\rm l}\sim 200\,\pc$, was also quite close
relative to the photometric distance inferred from the {\it HST}
photometry.
\citet{alcock01} suggested that more precise astrometric measurements with 
the new Advanced Camera for Surveys (ACS) on {\it HST} could determine 
the lens-source
relative parallax $\pi_\rel$, and so test the distance estimate and
hence effectively test the entire procedure for inferring both the
mass and distance of the lens.

In the meantime, \citet{gould04}, building on the work of \citet{smp},
discovered that microlens parallax measurements of relatively
short events (with timescales $t_\e\la {\rm yr}/2\pi$) are subject
to a four-fold degeneracy composed of two two-fold ambiguities.
While one these ambiguities (the so-called ``constant-acceleration''
degeneracy) has only a very small effect on the event
parameters, the other (the ``jerk-parallax'' degeneracy) can affect 
$\bpi_\e$ by quite a lot.  \citet{gould04} showed that the alternate
jerk-parallax solution for MACHO-LMC-5 had a somewhat larger mass
and a much larger distance, $D_{\rm l}\sim 450\,\pc$.

Most recently, \citet{drake} have carried out the ACS measurements
advocated by \citet{alcock01}, and these have yielded both a more
precise measurement of $\bmu_\rel$ and a new measurement of $\pi_\rel$.
\citet{drake} were able to conduct two consistency checks.  First,
they found that their trigonometric parallax measurement was 
consistent with the lens distance derived by \citet{gould04} for his
alternate ($D_{\rm l}\sim 450\,\pc$) solution.  Second, they found that the
direction vector $\bmu_\rel/\mu_\rel$ determined from astrometry
was consistent with the direction vector $\bpi_\e/\pi_\e$ found by
\citet{gould04} for this alternate 
solution.

In fact, the proper motion derived from the WFPC2 observations proved 
very accurate and agreed to within $0^\circ \hskip-2pt .2$ with the direction
from the new ACS 
measurements (both in 2000 celestial
coordinates).  However, \citet{alcock01} reported these results only
in ecliptic coordinates, and they made a $14^\circ$ degree error when
they translated from celestial to ecliptic.  This transcription
error significantly affected the mass and distance estimates of both 
\citet{alcock01} and \citet{gould04} when they used the direction of
proper motion as a constraint in their solutions.  The correction
of this transformation error, by itself, resolves the most puzzling
aspects of the solutions obtained by \citet{alcock01} and \citet{gould04}.
If the correct direction had been incorporated into the fits, both solutions
would have moved to within $1\,\sigma$ of the hydrogen-burning limit.  The
\citet{gould04} solution would have moved to $\sim 530\,\pc$ and so would
have been in better agreement with the photometric parallax, although the
\citet{alcock01} solution would have actually moved to an even shorter 
distance, $\sim 160\,\pc$.

\subsection{ACS Astrometry
\label{sec:acs}}

In this paper, we will make extensive use of the astrometric measurements
of \citet{drake}, sometimes combining them with other real and/or hypothetical
data.  To do so, we must fit to the original astrometric data.  We find that
when we fit these data alone, the results differ very slightly (much less
than $1\,\sigma$) from the results reported by \citet{drake}.  For
consistency, we always use the values from our own fits, 
although the difference
has no practical impact on any of the derived results.  In fitting the
\cite{drake} data, we exclude the 2002 F814 point as they also did
(A.\ Drake 2004, private communication).  This point is a significant
($3\,\sigma$) outlier, which contradicts a F606 measurement taken
at almost exactly the same time.  
%Systematic distortions
%in the astrometric reductions of {\it either} the F814 or F606 data
%would cause an offset between the F814 and F606 measurements.  There
%is no way to tell from the data which filter would have been affected.
%However, the parallax measurement depends primarily on a comparison
%of two F606 points taken six months apart.  If the distortion field
%is a high power (e.g. $\nu=3$) of the separation, then the relative
%error in the offset between these two measurements is smaller by
%a factor $(12\,\rm mas/210\,mas)^\nu$ than the relative error
%in the source-lens separation that is giving rise to the discrepancy
%between the F606 and F814 measurements.  As this relative error is 
%only $\sim 0.8\,$mas, the effect on the parallax measurement would
%be negligible, even if the source of the discrepancy lay in the
%F606 data.  On the other hand, if this discrepancy were due to a
%distortion in the F606 data, then there would be a systematic error induced
%in the proper motion of $\sim 0.8\,\rm mas/(9\,yr)\sim 90\,\mu as\,yr^{-1}$.
%While four times larger than the formal error in the proper motion,
%this is still far too small to have any practical effect on the results
%reported in this paper.  Hence, while the discrepancy between the F606 and
%F814 points is somewhat unsettling, elimination of the F814 is clearly
%the safest solution.

Our fit yields the following parameter estimates:
relative parallax $\pi_\rel = 1.780\pm 0.185\,$mas;
relative proper motion $\mu_\rel = 21.381\pm 0.022\,\rm mas\, yr^{-1}$;
proper motion components 
$\mu_{\rm rel,East} =17.547\pm 0.029\,\rm mas\, yr^{-1}$;
and $\mu_{\rm rel,North} = -12.217\pm 0.022\,\rm mas\, yr^{-1}$; and
position angle $\phi=124^\circ\hskip-2pt .85 \pm 0^\circ\hskip-2pt .08$.

As noted by \citet{drake}, the residuals for the WFPC2 point are quite small
compared to their reported errors, with $\Delta\chi^2=0.11$ for 2 degrees
of freedom (dof).  This may imply that the errors were overestimated by
\citet{alcock01}.  However, since there is a 10\% probability of
having such low residuals by chance, no definite conclusions can be drawn
regarding a possible overestimation of the error bars.

\section{Sources and Consistency of Data
\label{sec:consistdata}}

	In this paper, we will draw together four sources of
data to measure the mass of MACHO-LMC-5.  We first summarize these
sources, then discuss a series of tests that we have carried out to 
determine whether they are consistent with each other.  Only
after these tests are successfully concluded do we combine the
data.

\subsection{Data Sources
\label{sec:sources}}

The primary data set is the original MACHO SoDoPHOT pipeline photometry
of the event, which occurred in 1993.  These data have already
been analyzed by \citet{alcock01} and \citet{gould04}.  They consist
of 352 points in the non-standard MACHO red filter (hereafter $R_M$)
and 265 points in the non-standard MACHO blue filter (hereafter $V_M$).
We slightly deviate from previous authors by recursively removing
outliers and renormalizing the errors so as to enforce $\chi^2$
per degree of freedom (dof) equal to unity in each bandpass
separately.  We repeat this procedure until all $3.5\,\sigma$
outliers are removed.  This removes three $R_M$ points and one
$V_M$ point, all greater than $3.9\,\sigma$.  The next largest deviation is
at $3\,\sigma$, but with more than 600 points, such a deviation is 
consistent with Gaussian statistics and so cannot be considered an
outlier.  The error renormalization factors are 0.79 in $R_M$ and 0.81
in $V_M$.

\citet{gould04} had argued against blindly applying this renormalization
procedure because the mass determination is dominated by the relatively
small number of points during the event, while $\chi^2$/dof is dominated
by the much larger number of baseline points taken over several years.
However, we find that $\chi^2$/dof is similar for both these subsets
and therefore proceed with the renormalization described above.
Of the four eliminated points, only one is during the event, the
$R_M$ point at JD = 2490015.14.  From Figure 1 of \citet{gould04},
it can be seen that this point is a clear outlier with an abnormally
large error bar.

For isolated faint stars, SoDoPHOT (``Son of DoPHOT'')
reports only photon noise errors,
which of course cannot be overestimates.  However, in crowded fields,
SoDoPHOT follows DoPHOT \citep{dophot} in ``padding'' the photon
noise as it subtracts out surrounding stars.  This additional padding
is correct in some average sense, but may be an overestimate or
underestimate in individual cases.  Hence, for events of particular
interest, it is worthwhile to investigate these error bars more closely.

For Gaussian errors, removal of outliers greater than 
$\sigma_\max = 3.5$ artificially reduces $\chi^2$ by 
$\sim (2/\pi)^{1/2}\sigma_\max\exp(-\sigma_\max^2/2)\sim 0.6\%$
and so understates the size of the error bars by 0.3\%.
This difference has no practical effect on the results reported here.

The second data source is the {\it astrometric} measurement of the lens-source
separation made by analyzing epoch 1999 
{\it Hubble Space Telescope (HST)} Wide Field and Planetary Camera
(WFPC2) images originally obtained by \citet{alcockprelim}.
\citet{alcock01} and \citet{gould04} have previously combined the proper 
motion measurement, $\bmu_\rel$, derived from these data
with the above-mentioned lightcurve data to estimate the mass and distance of 
the lens.

The third data source is the {\it photometric} measurements of the source
brightness made from the same {\it HST} WFPC2 images.  These data
help constrain the lightcurve fit and thus tighten the errors on the
mass and distance measurements.  \citet{alcock01} made use of these
measurements in their constrained fit, but \citet{gould04} did not.

The final data source is the new astrometric measurements made by \citet{drake}
using ACS.  These have
yielded both an improved measurement of $\bmu_\rel$ and a new parallax
measurement, $\pi_\rel$.  Both the WFPC2 and ACS measurements are listed
in Table 1 of \citet{drake}. 

\subsection{Consistency Checks
\label{sec:consistency}}

\subsubsection{Source Color and Magnitude
\label{sec:colmag}}

We first wish to combine the original SoDoPHOT photometry of the event
with the flux measurement of the source made by \citet{alcock01}
after the event was over and the source was well separated from the lens.
The superb resolution of {\it HST} virtually ensures that all blended
light is removed from the source with the possible exception of a
wide-binary companion to the source, which we discuss below.
This {\it HST} source photometry can be compared with the source flux that 
is returned as a parameter by
the microlensing fit.  To do so, one must first translate the {\it HST}
photometry into the MACHO bands.  This can be done by directly
comparing the flux levels recorded by {\it HST} and SoDoPHOT for
an ensemble of other stars in the field.  While each of these is
blended in the MACHO images, the blending is equally likely to
contaminate the object or the ``sky'' determination.  So it should
introduce scatter but not a systematic bias.  This scatter is smaller
for brighter stars, but unfortunately the PC field is not big enough
to contain many bright stars.  Based on a comparison of 18 relatively
bright stars (and constraining the fits by the two color-color slopes,
$V-V_M\propto -0.20(V_M-R_M)$, $R-R_M\propto +0.18(V_M-R_M$)
reported by \citealt{machocal}), we find that the {\it HST} data imply
\begin{equation}
%f_{{\rm s},V} = 23.20 \pm 1.30,\quad
%f_{{\rm s},R} = 23.58 \pm 1.20,\quad
%{f_{{\rm s},R}\over f_{s,V}} = 1.017 \pm 0.018\qquad (HST),
f_{{\rm s},V} = 23.27 \pm 1.34,\quad
f_{{\rm s},R} = 23.84 \pm 1.23,\quad
{f_{{\rm s},R}\over f_{s,V}} = 1.024 \pm 0.016\qquad (HST),
\label{eqn:hstcolors}
\end{equation}
where the errors and covariances are derived by enforcing $\chi^2/$dof 
in the fit.  The error in the color calibration is substantially smaller
than the errors in the flux calibrations because the latter are highly
correlated, with correlation coefficient $\rho=0.962$.
These results may be compared to the source flux levels derived from
the  fit to the SoDoPHOT data alone,
\begin{equation}
f_{{\rm s},V} = 28.97 \pm 3.99,\quad
f_{{\rm s},R} = 29.50 \pm 4.16,\quad
{f_{{\rm s},R}\over f_{{\rm s},V}} 
= 1.018 \pm 0.018\qquad ({\rm SoDoPHOT\ lightcurve}).
\label{eqn:sodophotcolors}
\end{equation}
The two determinations differ by $\sim 1.3\,\sigma$ in each 
of the two (highly correlated) bands
separately and by $\sim 0.3\,\sigma$ in the color.  Since
the two photometric measurements are consistent, they can be combined.
We then find that the $\chi^2$ minimum increases
by 2.2 for two additional dof, confirming the consistency of the
two pairs of measurements.

The one possible caveat is that the microlensing fit gives the flux
of the source that was magnified during the event while the {\it HST}
measurement gives all the flux from stars within about 100 mas of the
source center, corresponding to about 5000 AU at the distance of the
Large Magellanic Cloud (LMC) \citep{alves04}.  
Significant sources of light that
lay beyond 50 mas would have shown up in subsequent ACS images.
Now, as we will show below,
the angular Einstein radius is $\theta_\e \sim 1$ mas.  Any significant light
source within about $(1/3)\theta_\e$ of the source would have betrayed
itself during the event.  This still leaves the possibility that
the source has a binary companion between 17 and 2500 AU.  While we cannot 
rule out such a possibility, there are several lines of argument against it.
First, the discrepancy (eqs.~[\ref{eqn:hstcolors}] and 
[\ref{eqn:sodophotcolors}])
between the two measurements (though not statistically
significant) is of the {\it wrong sign} to be accounted for by a binary
companion.
Second, the good agreement in the colors shows that any companion must
be either of nearly the same color as the source or quite faint.
If the former, the two stars should also be of roughly the same magnitude, in
which case one would expect the discrepancy to be much larger than observed
(and, again, in the opposite direction).  Third, to produce $\sim 10\%$
or more of the light, the companion mass would have to be $>70\%$ of the 
primary.  If LMC binaries are similar to those studied by
\citet{dm91} in the Galactic disk, the fraction of stars with companions
in the required mass and separation ranges is only $\sim 7\%$.  Hence,
it is unlikely, though not impossible, that such a companion exists
and is significantly corrupting the measurement.

\subsubsection{Direction of Motion
\label{sec:direction}}

\begin{figure}
\plotone{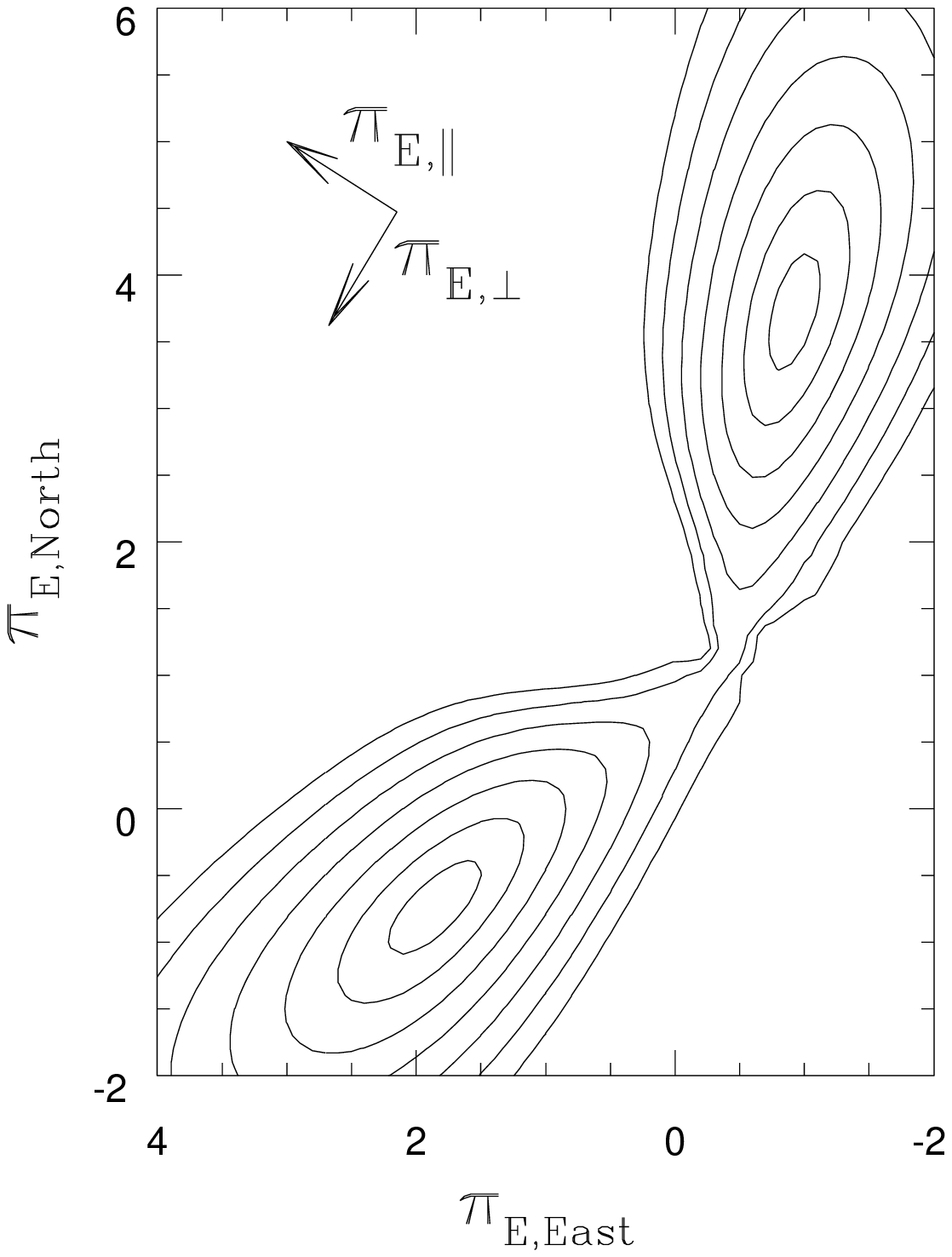}
\caption{Likelihood contours in the $\bpi_\e$ plane shown at
$\Delta\chi^2=1$, 4, 9, 16, 25, 36, and 49 relative to the minimum.
This should be compared to Figure 3 of \citet{gould04}.  The
errors here are smaller, partly because of error renormalization
and partly because of the addition of {\it HST} baseline 
photometry of the source from \citet{alcock01}.
\label{fig:piecontours}}
\end{figure}

We therefore begin by incorporating the WFPC2 photometric measurements
into the lightcurve fit, taking account of both their errors and
covariances.
The resulting contour plot for the vector parallax $\bpi_\e$ (in the geocentric
frame) is shown in Figure~\ref{fig:piecontours}.  This figure should
be compared to Figure 1 from \citet{gould04}.  Each of the two minima
is consistent between the two figures at the $1\,\sigma$ level.  This
is to be expected, since the additional photometric data are
consistent with those used by \citet{gould04}.  However, the errors
are substantially smaller, both because the photometric errors have
been renormalized by a factor $\sim 0.8$ and because of the additional
higher-precision {\it HST} baseline photometry of the source.

We can now ask whether the direction of motion (in the heliocentric frame)
implied by each of these lightcurve ($\phi_{\rm lc}$) 
solutions is consistent with the direction of proper motion ($\phi_{\rm ast}$) 
that we derive from the {\it HST} ACS and WFPC2 data of \citet{drake}.
For the southeast solution, which the \citet{drake} parallax measurement
demonstrates to be the correct one, the comparison yields
\begin{equation}
\phi_{\rm ast} = 124^\circ\hskip-2pt .85 \pm 0^\circ\hskip-2pt .08
\qquad
\phi_{\rm lc} = 132^\circ\hskip-2pt 
.3^{+ 3^\circ\hskip-2pt .4}_{-5^\circ\hskip-2pt .3}.
\label{eqn:phicomp}
\end{equation}
Hence, the photometrically and astrometrically determined directions
are consistent at about the $1.3\,\sigma$ level.  
The other (northwest) solution has a direction
$\phi_{\rm lc} = 136^\circ\hskip-2pt 
.5^{+ 2^\circ\hskip-2pt .2}_{-2^\circ\hskip-2pt .5}$. However,
because the $\chi^2$ surface deviates strongly from a parabola,
the discrepancy with the proper-motion data is only at about the $2.8\,\sigma$
level.

\subsubsection{Parallax
\label{sec:parallax}}

\begin{figure}
\plotone{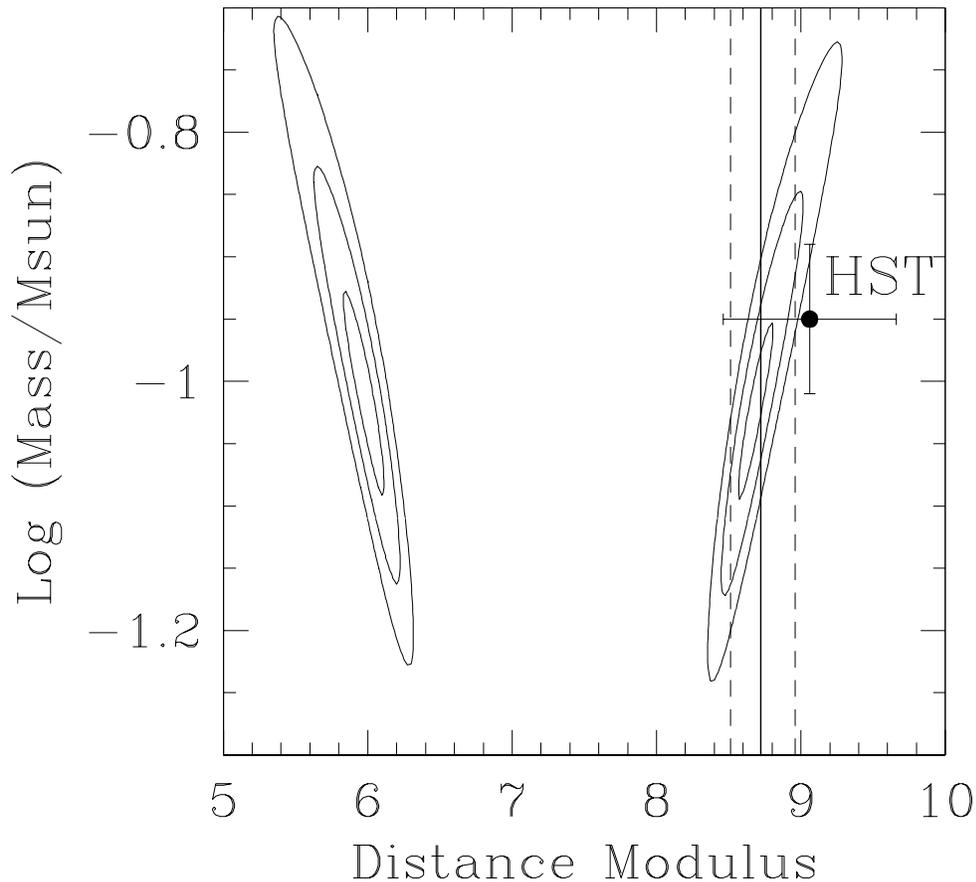}
\caption{Likelihood contours ($\Delta\chi^2=1,4,9$)
for the log mass $M$ and distance modulus
$(M-m)_0$ of MACHO-LMC-5
based on the lightcurve of the event, the source-flux measurement
of \citet{alcock01}, and constrained by the
proper-motion measurement of \citet{drake}.  Each set of contours
is shown relative to its own minimum.  The left-hand minimum
is actually higher than the one at the right by $\Delta\chi^2=6$.
The vertical
lines show the best-fit distance modulus and $1\,\sigma$ confidence interval
derived from the trig parallax 
measurements of \citet{drake}, while the point with error bars 
shows $\log M$ and $(M-m)_0$ as determined photometrically from
{\it HST} WFPC2 data by \citet{alcock01}.  
Note that the photometric (contours)
and astrometric (vertical lines) determinations of $(M-m)_0$ are
in agreement at better than $1\,\sigma$.
\label{fig:massdist1}}
\end{figure}

Since the astrometrically determined direction of motion is
consistent with the value derived from the lightcurve (at least
for the southeast solution), we combine the astrometric and
photometric data.  As explained by \citet{alcock01} and \citet{gould04},
this permits a full solution for the event, including the mass
$M$ of the lens and lens-source relative parallax $\pi_\rel$.
We plot the result in the [$\log M$,$(M-m)_0$] plane, where
$(M-m)_0$ is the {\it lens} distance modulus, i.e., corresponding
to $\pi_{\rm l} = \pi_\rel + \pi_{\rm s}$ and where we have
adopted $\pi_{\rm s} = 20\,\muas$ for the source, which resides in the LMC.  
Figure~\ref{fig:massdist1}
shows the resulting likelihood contours.  Note that each set of
contours is offset from its respective minimum.  The minimum
of the short-distance (northwest) solution is actually higher by
$\Delta\chi^2=6$, which is a reflection of the mild direction
discrepancy found for this solution in \S~\ref{sec:direction}.

Also shown in Figure~\ref{fig:massdist1} is the best fit and $1\,\sigma$
error bar for the parallax determination based on the astrometric
data of \citet{drake}, $\pi_\rel = 1.780\pm 0.185\,$mas.
This corresponds to a distance modulus
\begin{equation}
(M-m)_0 = 8.72^{+0.24}_{-0.21}\qquad (\rm trig\ parallax).
% 8.96 8.51
\label{eqn:trigpi}
\end{equation}
To make an algebraic comparison, we fit the $\Delta\chi^2<1$ region of
the rightward contours to a parabola and find,
\begin{equation}
\log M = -1.023 \pm 0.084,\quad (M-m)_0 = 8.683\pm 0.144, \quad \rho=0.921,
\qquad (\rm lc+\bmu_\rel )
\label{eqn:errellipse1}
\end{equation}
where $\rho$ is the correlation coefficient.  Hence, the lens distance
derived from the lightcurve/proper-motion analysis is consistent
with the trig parallax at the $1\,\sigma$ level.

\section{Mass, Distance, and Velocity of Lens
\label{sec:massdist}}

Since the two measurements are consistent, we combine them.  The
results are shown in Figure~\ref{fig:massdist2} and can be represented
algebraically by,
\begin{equation}
\log M = -1.013 \pm 0.073,\quad (M-m)_0 = 8.702\pm 0.124, \quad \rho=0.896.
\qquad (\rm lc+\bmu_\rel+\pi_\rel )
\label{eqn:errellipse2}
\end{equation}
These figures correspond to a best-fit mass and distance
\begin{equation}
M=0.097\pm 0.016\,M_\odot,\qquad D_{\rm l}=550\pm 30\,\pc.
\label{eqn:massdistvals}
\end{equation}
  This best fit has
$\chi^2=605.39$ compared to $\chi^2=601.68$ for the lightcurve alone.  
That is, $\Delta\chi^2=3.71$ for 5 additional dof.

Also shown in Figures~\ref{fig:massdist1}
and \ref{fig:massdist2} are the estimates of the mass
and distance of the lens as derived by \citet{alcock01} from
photometric {\it HST} data.  These are consistent with the microlensing
measurement at the $1\,\sigma$ level.

The $(U,V,W)$ velocities of the lens toward the Galactic center, the
direction of Galactic rotation, and the north Galactic pole are
\begin{equation}
U=43.6\pm 1.9\,\kms,\qquad
V=-60.8\pm 8.3\,\kms,\qquad
U=26.6\pm 5.7\,\kms,
\label{eqn:uvw}
\end{equation}
with correlation coefficients $\rho_{UV}=-0.90$, $\rho_{UW}=-0.74$, 
$\rho_{VW}=0.93$, where we have taken account of the source motion
\citep{vdm} and the motion of the Sun relative to the
local standard of rest \citep{db}.
The uncertainties are dominated by the error in the radial-velocity
measurement, $v_r=49\pm 10\,\kms$ \citep{alcock01}, and this fact
accounts for the high correlation coefficients.

Finally, for reference we note that, from equations~(\ref{eqn:thetadef})
and (\ref{eqn:piedef}), these determinations of the mass and distance imply
\begin{equation}
\theta_\e = 1.19\pm 0.07\,{\rm mas},\qquad
\pi_\e = 1.51\pm 0.17,
\label{eqn:thetaepie}
\end{equation}
where we have taken account of the correlations in determining
the errors.

\section{Constraints on Binarity
\label{sec:binarity}}

An important application of microlensing mass measurements is
the opportunity they afford to confront theoretical models
that attempt to predict the masses of stars from their spectroscopic and
photometric properties.  Crucial to such a comparison is the
determination that the ``star'' is in fact a single object and
not a close stellar binary or a binary composed of a star and
a brown dwarf.  In the former case, the photometric properties
would be a composite and in both cases the microlens mass would
not be the mass of the star dominating the light.  It is equally
crucial that the luminous star whose visible properties are
being measured is actually the lens whose mass was measured during
the microlensing event, as opposed to a luminous companion of
a dark object (e.g.\ a brown dwarf) that generated the microlensing
event.  In this section we investigate how well both of these concerns
can be addressed with current and/or future data.

\subsection{Limits on Close Binaries
\label{sec:close}}

A close binary approximates a \citet{cr1,cr2} lens with sheer
$\gamma=[d/(q^{1/2}+q^{-1/2})]^2$, where $d\theta_\e$ is the
angular separation of the two components and $q$ is their 
mass ratio \citep{dominik99,albrow02}. These have caustics of full
angular width $4\gamma\theta_\e$, which, if traversed, would give rise 
to obvious
deviations from a point-lens lightcurve.  For a continuously sampled
lightcurve, the best way to avoid the caustic is for the source
to pass at an angle of $45^\circ$ relative to the binary axis.
In this case, the caustic is just barely nicked if $u_0=2^{1/2}\gamma$,
where $u$ is the lens-source separation in units of $\theta_\e$
and $u_0$ is its minimum value (i.e., impact parameter).  For
MACHO-LMC-5, the sampling is far from uniform, and the tightest 
simple
constraint is obtained from the highest point, which comes almost
exactly one day after the peak and therefore is at 
$u=u_{\rm high}=0.029$.  The largest circle that one can inscribe in
the central caustic has radius $u\sim\gamma$.  From this we derive,
\begin{equation}
\gamma = {d^2\over (q^{1/2}+q^{-1/2})^2} < u_{\rm high}=0.029,
\label{eqn:gammalim}
\end{equation}
since otherwise this point would have landed in the caustic and so would have
been much more magnified than it actually is.  Since the lens is much
closer than the source, the Einstein radius is essentially equal
to the projected Einstein radius, $r_\e=\tilde r_\e = \au/\pi_\e = 0.66\,\au$.
Hence the above limit can be expressed in terms of the projected separation
between the binary components, $r_\perp \equiv d r_\e$,
\begin{equation}
r_\perp < 0.11(q^{1/2}+q^{-1/2})^2\,\au.
\label{eqn:rperplim}
\end{equation}
Unless we are very unfortunate to see a widely separated pair projected
along the line of sight, or unless the companion is of such low mass
as to be uninteresting, the putative companion would cause the source
to move by $10\,\kms$ or more.  This would in principle be detectable
by spectroscopic measurements.

While close binaries deviate most sharply from point lenses inside
their caustics, they do show significant deviations in the
surrounding regions as well (see Fig.~1 from \citealt{gg97}).
A detailed accounting of these deviations would strengthen the limit
in equation~(\ref{eqn:rperplim}) but, given the sparse sampling of MACHO-LMC-5,
the improvement would most likely be modest.

\subsection{Limits on Wide Binaries
\label{sec:wide}}

A similar argument places a limit on wide companions, which also
give rise to Chang-Refsdal caustics.  In this case, $\gamma= q d^{-2}$, so
\begin{equation}
r_\perp > 3.9\,q^{1/2}\,\au,
\label{eqn:rperplim2}
\end{equation}
corresponding to $7.2\,q^{1/2}$mas.

\subsection{Constraints on the Dark Lens Hypothesis
\label{sec:dark}}

As shown in Figures~\ref{fig:massdist1} and \ref{fig:massdist2},
the best-fit mass lies close to the hydrogen burning limit.
It is therefore possible in principle
that the lens is not actually the red star seen in the {\it HST}
images but rather an invisible brown-dwarf companion to it.
To what extent can this scenario be constrained by the available data?

As pointed out by \citet{drake}, even if one relaxes the assumption
that the red star and the microlensed source were coincident at the
time of the event, the remaining astrometric measurements ``point back''
to a relative offset $\Delta\btheta$
very close to zero at the peak of the event.
Specifically, we find,
\begin{equation}
\Delta\theta_{N} = -1.2\pm 7.7\,{\rm mas}
\qquad
\Delta\theta_{E} = -2.9\pm 7.5\,{\rm mas}.
\label{eqn:deltatheta1}
\end{equation}
As discussed by \citet{drake}, the smallness of these values relative 
to the errors most likely reflects an overestimation of the {\it HST}
WFPC2 errors by \citet{alcock01}.  However, to be
conservative, we ignore this possibility.  Somewhat stronger constraints
can be obtained by noting that in this relaxed solution, the parallax
error grows from 0.185 mas to 0.291 mas, and that this error is
fairly strongly correlated with $\Delta\btheta$.  However, the microlensing
analysis independently constrains the parallax to be $\pi_\rel=1.81\pm 0.12$
mas, and this constraint can be added into the fit.  We then find,
\begin{equation}
\Delta\theta_{N} = -0.8\pm 5.6\,{\rm mas}
\qquad
\Delta\theta_{E} = -2.5\pm 6.8\,{\rm mas}.
\label{eqn:deltatheta2}
\end{equation}
These results indicate that, at the $2\,\sigma$ level, the M dwarf must
have been within about 13 mas of the source at the time of the event.
On the other hand, from the argument in \S~\ref{sec:wide}, the M dwarf
could not have been too close to the source if it were not actually the
lens.  By hypothesis, the putative brown-dwarf lens is below the
hydrogen-burning limit while the M dwarf is above it, so $q>1$.
Hence, 
\begin{equation}
\Delta\theta = {r_\perp\over D_{\rm l}} > 7.2\,\rm mas.
\label{eqn:wide2}
\end{equation}

Equations~(\ref{eqn:deltatheta2}) and (\ref{eqn:wide2}) leave only
a fairly narrow range of allowed separations.  For face-on circular
orbits and for a total binary mass $M_{\rm tot}=0.2\,M_\odot$,
these limits correspond to a period range
\begin{equation}
18\,{\rm yrs} < P < 44\,{\rm yrs}\qquad \rm (allowed\ periods).
\label{eqn:periodallow}
\end{equation}
Even assuming a mass ratio $q=2$, the amplitude of the M dwarf motion
would be between 2.4 mas and 4.3 mas.  These amplitudes
are quite large relative to the $\sim 0.3\,$mas errors achieved for
single epoch {\it HST} ACS images.  Hence, in principle, this 
scenario could be much more tightly constrained by future observations.

\section{Future Confrontations
\label{sec:future}}

\begin{figure}
\plotone{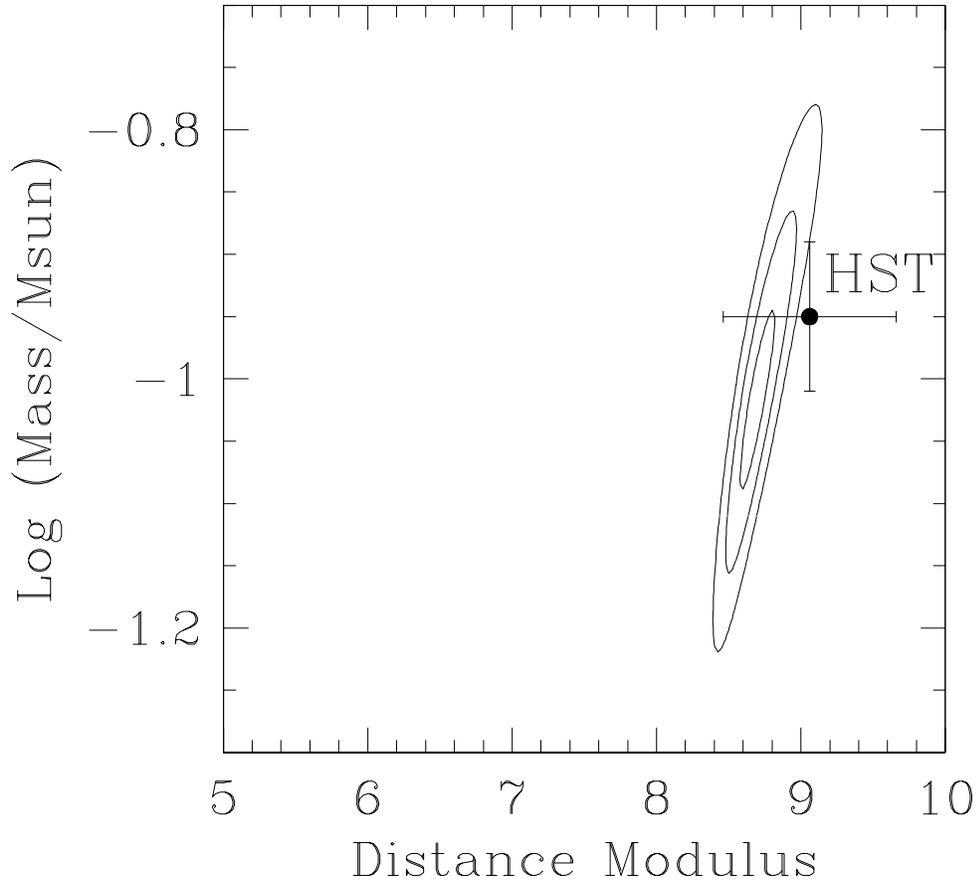}
\caption{Similar to Fig.~\ref{fig:massdist1} except that the 
microlensing/proper-motion determination has now been combined
with the lens-source relative parallax measurement of \citet{drake}.
The errors are 17\% in the mass and 6\% in the distance.  There
is good agreement with the mass and distance estimates based
on {\it HST} photometry, shown as a point with error bars.
\label{fig:massdist2}}
\end{figure}

	From Figure~\ref{fig:massdist2}, the photometrically derived mass
and distance of MACHO-LMC-5 are consistent with the values of these
properties derived by combining the astrometric and microlensing data.
The error bars for both determinations could be improved significantly
by obtaining additional data and by improving the analysis.

	On the photometric side, both the mass and the luminosity of the 
M dwarf are inferred from its color.  Apart from the error in measuring
this quantity, these inferences suffer from the intrinsic dispersions
of mass and luminosity at fixed color.  A substantial part of this
dispersion is due to metallicity.  \citet{drake} have argued that
the kinematic data are consistent with either a disk or a thick-disk
star, and therefore a range of metallicities of about 1 dex.  Hence,
spectroscopic determination of the M dwarf's metallicity would go
a long way toward shrinking the photometry-based mass/distance error bars.

	On the astrometric/microlensing side, there are three paths
to improvement.  First, of course, the distance determination
could be improved by a better trigonometric parallax measurement.
Moreover, because the mass and distance measurements
are highly correlated (see eq.~[\ref{eqn:errellipse2}]), a more
accurate distance would also improve the mass determination.
Unfortunately, significantly better parallax measurements will
not come cheaply.  From a comparison of 
equations~(\ref{eqn:trigpi}) and (\ref{eqn:errellipse1}), the
$185\,\muas$ error from the ACS astrometry is about 50\%
larger than the distance estimate achieved from microlensing 
(and the proper-motion measurement) alone.
A plausible target for a significant improvement would be
a $100\,\muas$, or better yet $50\,\muas$ measurement.  These
would yield mass determinations with fractional precisions
of 13\% and 10\%, respectively.  

Note that even if the distance were known
exactly, the microlens mass measurement error could only be reduced
to 7.5\%.

If it were only necessary to consider the statistical errors, 
such improvements could be achieved by multiplying the total length 
of ACS observations by 3 and 14 respectively.  However, systematic
errors may become important, and the discrepancy between the
F606 and F814 measurements implies that caution is warranted.

A parallax measurement by {\it SIM} might also prove feasible.
This seems impossible at first sight because the M dwarf
has $V=22.7$ whereas the magnitude limit of {\it SIM} is often
said to be $V<20$.  However, what fundamentally limits {\it SIM}
at faint magnitudes is the number of sky photons that enter
its $1''$ radius stop.  If we ignore this sky noise for the moment,
a $50\,\muas$ measurement at $V=22.7$ would require
an observation of only about 1 hour.  An additional 
30-minute observation of the $V=21$ source would yield a $30\,\muas$
measurement, for a combined error in the relative offset of $60\,\muas$.
As in the case of the ACS measurements, only two epochs 
(in two orientations) would be
required because it is known from the microlensing event that the
two stars were virtually coincident in 1993. We find that a total
of 4 pairs of observations (each pair totaling 90 minutes)
would yield
a relative parallax error of $42\,\muas$.  The sky
is $V\sim 22.7\,\rm mag\,arcsec^{-2}$  (so $V\sim 21.5$ inside the {\it SIM}
stop), which would mean that the observation time for the lens (but not
the source) would have to be roughly tripled relative to the naive
estimates.  However,
{\it SIM} has a very broad bandpass and the M dwarf is very
red, which may imply that much of the astrometric signal will come in
well above the sky.  Hence, the required duration of the exposures
cannot be properly estimated until the details of {\it SIM}'s
throughput are worked out in greater detail.  In any event,
it appears that {\it SIM} could achieve a substantial improvement in 
the parallax measurement without prohibitive observing time.

The second potential path would be to obtain better photometry of the source.
Recall that the {\it HST} flux measurement error was dominated
by the problem of aligning the WFPC2 and SoDoPHOT photometry, which
was exacerbated by the small number and faint flux levels of 
stars in the small PC chip.  Since the source and lens are now
well separated, one could image them using the much larger ACS
camera and so align the photometry using a large number of
relatively bright stars.  If the photometric error were reduced
from the present 5.5\% to 2\%, then the mass error would be
reduced from 17\% to 15\%.  If this were combined with
a $50\,\muas$ parallax measurement, the mass error would be reduced
from the above-mentioned 10\% to 7\%, while perfect knowledge of
the distance would, under these circumstances, reduce the mass
error to 4\%.

The third path would be improved microlensing data reductions.
All the microlensing analysis has been conducted on the basis of the
original MACHO SoDoPHOT pipeline photometry.  This pipeline produced
$10^{10}$ measurements of very high quality, but with modern
image-subtraction routines, it may be possible to do better.
However, since the SoDoPHOT errors have been renormalized
(see \S~\ref{sec:sources}) some of this improvement has already been achieved.

%\begin{equation}
%\label{eqn:}
%\end{equation}

\acknowledgments
A.G. was supported by grant AST 02-01266 from the NSF.
D.P.B was supported by grants AST-0206189 from the NSF and NAG5-13042
from NASA.

\clearpage

\clearpage

%\begin{figure}
%\plotone{f3.ps}
%\caption{
%\label{fig:}}
%\end{figure}

\end{document}